\documentstyle[aps,epsf]{revtex}

\begin{document}

\twocolumn[
\hsize\textwidth\columnwidth\hsize\csname@twocolumnfalse\endcsname\title
{Statistical properties of the attendance time series in the minority game}

\author{Dafang Zheng$^{a}$ \and\ Bing-Hong Wang$^{b,c}$}

\address{$^{a}$ Department of Applied Physics, 
South China University of Technology,\\%
Guangzhou 510641, P.R. China, E-mail: phdzheng@scut.edu.cn\\
$^{b}$ Center of Nonlinear Science and Department of Modern Physics, \\
University of Science and Technology of China, Hefei 230026, P.R.China\\
$^{c}$ Department of Physics, The Chinese University of Hong Kong, \\
Shatin, New Territories, Hong Kong, E-mail: bhwang@phy.cuhk.edu.hk}
\maketitle

\begin{abstract}
We study the statistical properties of the attendance time series
corresponding to the number of agents making a particular decision in the
minority game (MG). We focus on the analysis of the probability distribution
and the autocorrelation function of the attendance over a time interval in
the efficient phase of the game. In this regime both the probability
distribution and the autocorrelation function are shown to have similar
behaviour for time differences corresponding to multiples of $2\cdot 2^{m}$,
which is twice the number of possible history bit strings in a MG with
agents making decisions based on the most recent $m$ outcomes of the game. \\%
\end{abstract}

]

The minority game (MG), introduced by Challet and Zheng\cite{challet},
represents a simplified version of the El Farol bar-attendance problem
\cite{arthur1,arthur2,bar}
proposed by Arthur. It gives perhaps the simplest
model of a complex adaptive system \cite{holland} in which agents of similar
capability are competing to be in the minority based on some globally shared
information. The model may be useful in investigating many of the features
observed in financial markets\cite{farmer,stanley}. In its basic form, the MG
describes a system (market) in which an odd number $N$ of agents are allowed
to make two possible choices in each turn. At each step, agents have to
choose either to be in side $0$ (buying) or side $1$ (selling). After every
agent has independently chosen a side, the side with fewer agents (the
minority side) is identified as the winning side. The ``output" of each time
step can be represented by a single binary digit: $0$ for side $0$ winning,
and $1$ for side $1$ winning. The record of the outputs for the last $m$
steps is the only information given to all agents based on which each agent
is to decide which side to take in the next time step. Therefore, there are
a total of $2^{m}$ possible history $m$-bit strings. The whole strategy
space thus consists of $2^{2^{m}}$ strategies. The number of strategies
grows rapidly with $m$. Fortunately, it has been pointed out \cite
{challet1,savit} that a reduced strategy space consisting of $2^{m}$ pairs
of mutually anti-correlated strategies, thus having a total of $2\cdot 2^{m}$
strategies, is sufficient to represent the full strategy space. In the
beginning of the game, each agent randomly picks $s$ strategies, with
repetitions allowed. After each step, each agent assigns (deducts) one
(virtual) point to each of his strategies which would have predicted the
correct (incorrect) output. At a given time step, each agent decides based
on the prediction of the most successful strategy in his bag of $s$
strategies. The past history thus creates a feedback mechanism that leads to
adaptivity in the population.

The MG and some of its variations have been the subject of much recent
attention\cite{challet2,EMG,TMG1,TMG2,TMG3,MEMG1,MEMG2}. One of the most
important features in the MG is that the average standard deviation $\sigma$
of the attendance time series in any one of the two sides over different
runs in computer simulations could drop to values better than the case in
which agents' decisions are made randomly. For small $s$, $\sigma$ has a
minimum as a function of $m$\cite{challet,savit}. If the number of
strategies in the reduced strategy space is small compared to the total
number of agents, i.e., $2\cdot 2^{m} \ll N\cdot s$, many agents tend to use
the same strategy at any given time step and hence make the same decision.
They form a ``crowd" that leads to the large standard deviation in the
statistics of the attendance. This regime is refereed to as the efficient
phase since the agents cannot make use of the information hidden in the
history bit-string as hidden information lies in strings longer than $m$ 
\cite{savit,deCara}. In the large $m$ limit, the strategy pool is much
larger than the number of strategies actually in play. The agents are
essentially making independent decisions randomly, leading to $\sigma \sim 
\sqrt{N}/2$ in this random-coin-toss limit. The minimum in $\sigma$ occurs
near $2\cdot 2^{m} \sim N\cdot s$ in which there are about the same numbers
of agents playing a strategy (forming a crowd) and its anticorrelated
partner (forming an anti-crowd) \cite{crowd1,crowd2}. It has been found that
the MG has very rich statistical features in the efficient regime\cite
{deCara}.

In this work, we focus on the changes in the attendance time series of the
MG in the efficient regime. Analogous to the statistical analysis of the
log-return in financial time series, we study the statistics of the
logarithmic changes in the attendance time series of MG. We also study the
autocorrelation function of the logarithmic change time series.

For a given time series of the number of agents $N_{0}(t)$ making a
particular choice, say taking side $0$, in the MG, we construct the time
series of the successive differences of the natural logarithm of the number
of attendance as \cite{stanley1} 
\begin{equation}
G_{\Delta t} (t) = \ln N_{0}(t + \Delta t) - \ln N_{0}(t),
\end{equation}
where $\Delta t$ is the sampling time interval. Such time series of
logarithmic changes in prices have been extensively studied
in the context of econophysics \cite{stanley2}.
A time series can be constructed for a given $\Delta t$. First we look
into the statistics of the values of $G_{\Delta t}(t)$ for given $\Delta t$.
Figure 1 shows the probability distribution $P$ of $G_{\Delta t}(t)$ for $%
\Delta t = 1$, $5$, and $8$ for the case of $N=1001$, $s=2$ and $m=2$. All
results are averaged over $32$ independent runs. Note that the probability
distributions for $\Delta t = 1 $ and $\Delta t=5$ show very clear
multi-peak structures, while for $\Delta t =8$, the peak corresponding to
zero-return dominates. Our results for $\Delta t =1$ are consistent with
those in Ref.\cite{deCara}. In Ref.\cite{deCara}, a plot of $N_{0}(t+1)$
against $N_{0}(t)$ was constructed. It was found that the plot consists of
patches reminiscence of attractors in a map. The existence of patches
implies the existence of dominant values in $G_{\Delta t=1}(t)$ as shown in
Fig.1(a). It is related to the virtual point assignments through which the
performance of the strategies are rated and the limited adaptability of the
agents imposed by a small value of $s$. The present work hence represents an
extension of the study in Ref.\cite{deCara} to other values of $\Delta t$
and investigate the interesting pheneomena of a periodicity of $2\cdot 2^{m}$
in time series in MG. The results in Fig.1(b) imply that similar patches
also appear in the $N_{0}(t+5)$ against $N_{0}(t)$ plot, which is given in
Fig.2(a). The plot is averaged over different runs and different return
maps with different initial time $t$. For $\Delta t = 8$, the sharp peak for
zero returns implies the formation of
patches concentrated among the diagonal in a plot of $%
N_{0}(t+8)$ against $N_{0}(t)$ plot (see Fig. 2(b)). We note that the
separate return map plotting non-overlapping data given by $N_{0}(t)$, $%
N_{0}(t+8)$, $N_{0}(t+16)$, $\cdots$ cover only a few patches in Fig.2(b)
for given initial value $t$.
After investigating the
distribution for higher values of $\Delta t$, we found that the probability
distribution corresponding to a given $\Delta t$ is identical to that
corresponding to $\Delta t + 2\cdot 2^{m}$. Thus the probability
distributions show identical features for different values of $\Delta t$
which are separated by multiples of $2\cdot 2^{m}$. For $m=2$, it implies
that $P$ for $\Delta t = 9$ is identical to that in Fig.1 (a) for $\Delta t=1
$. Similar properties are also observed for $m=3$ and other values of $m$
in the efficient phase for small values of $s$. 
Figure 3 shows the probability of zero
return $R$, which corresponds to the value of the zero-return peak in Fig.1,
as a function of $\Delta t$ for $m=2$ and $m=3$. It is observed that values
for different values of $\Delta t$ which are multiples of $2\cdot 2^{m}$
apart are nearly identical. The sharp peaks at $\Delta t = n \cdot 2\cdot
2^{m}$, where $n$ is a positive integer, reflect the dominant peak in the
probability distribution for $\Delta t = 8$ (see Fig.1(c)). Similarly for $%
m=3$, sharp peaks appear at $\Delta t$ equal to multiples of $2\cdot 2^{m}
= 16$ (see Fig.3(b)).

It is also interesting to look at the autocorrelation function of the
log-return time series. For $\Delta t = 1$, the autocorrelation $C(k)$ of $%
G_{1}(t)$ is 
\begin{equation}
C(k) = \frac{<G_{1}(t) G_{1}(t+k)> - <G_{1}(t)>^{2}} {<G_{1}^{2}(t)> -
<G_{1}(t)>^{2}},
\end{equation}
where $k$ is called the time lag. The averages $<\cdots>$ are taken over the
time series. Figure 4 shows the autocorrelation function for $m=2$ and $m=3$%
. Again, periodicity of $2 \cdot 2^{m}$ is evidence.

The statistics of the time series $G_{\Delta t} (t)$ of logarithmic changes
in the number of agents making a particular decision in the MG show
interesting statistically periodic features with period $T = 2\cdot 2^{m}$
in the efficient regime. The period $T=
2\cdot 2^{m}$ turns out to be twice the number of possible history
bit-strings in a game of given $m$. The observation can be understood
qualitatively as the system goes back to a similar situation only after
multiples of $2 \cdot 2^{m}$ time steps\cite{savit,rodgers}. Imagine in
the beginning of the game, each agent when encountering a particular history
will choose a strategy at random in making the decision as the virtual
points are all equal in the beginning. After the outcome is made known, the
strategies which gave the correct prediction gain one point. Assuming the
game visits each possible $m$-bit history with equal probability, the
particular history bit-string will be encountered about $2^{m}$ time steps
later as there are $2^{m}$ possible history bit-strings. At this time, the
agents holding the strategies which predicted correctly in the previous
occurrence of the history would make the same decision and due to
overcrowding they will lose. Virtual points will then be deducted from these
strategies and virtual points will be assigned to the other strategies.
After about another $2^{m}$ time steps, the history will be encountered
again. This time all the startegies have similar virtual points and the
situation is back to that in the beginning of the game, hence leading to the
observation of similar behaviour in the statistics of $G_{\Delta t}(t)$ for $%
\Delta t$ differs by multiples of $2\cdot 2^{m}$. The periodic behaviour
hence is a result of the interplay between the memory $m$ of the agents and
the limited adaptability when $s$ is small.  Difference in the
behaviour of the game for a particular history having occurred an odd number
and an even number of times has been discussed within the context
of strategy selections in MG \cite{rodgers}.

The behaviour is also related to the features in the statistics of the
occurrence of bit-strings of various lengths in the history, which is a
binary series. Close observation on the occurrence of
history bit-strings of length $m$
reveals the same periodic behaviour in that the history occurrence almost
repeats itself every $2 \cdot 2^{m}$ turns 
when $s$ is small, i.e. when the agents have
limited adaptability. The periodicity is,
however, too long for the agents with memory $m$ to spot it. For larger
values of $s$, the periodic behaviour become less obvious. Similarly, it has
been pointed out that \cite{savit} for MG in the efficient regime,
bit-strings of length $m+1$ or shorter occurs evenly while bit-strings of
length $k > m+1$ occurs unevenly. For $m=2$, $s=2$, and $N=1001$, uneven
bit-strings distribution arises for $k \geq 4$. Interestingly, for $k=
2\cdot 2^m = 8$, it is found that the string $11101000$ and $11100010$  
and their permutations
dominate the bit-string distribution. It is consistent with the even
distribution for $k \leq 3$ as these particular 8-bit strings contain all the
possible 3-bit strings.  In fact, given that the bit-strings must 
occur uniformly upto $k_{max}=3$ and there is some periodicity in
longer bit-string, the shortest length that is compatible with
the requirement is $2^{k_{max}}=8$.
For $k_{max}=3$, the total number of
different $3$-bit history strings is $8$.  The bit-string with the shortest
length that contains all the eight $3$-bit strings once must be of length $8$.   
This result reveals one more time the embedded
periodicity of $2\cdot 2^m$ in the MG in the efficient regime when the
agents' adaptability is limited. For large $m$, the strategy pool is huge
and the game approaches the random coin-toss limit in which the agents are
efficiently choosing a decision at random in each turn. In this case, the
multi-peak structure in the probability distribution of $G_{\Delta t}(t)$
will no longer persist.

In summary, we have studied the statistics of the time series of the
successive differences of the natural logarithm of the number of attendance
in MG. Interesting behaviour with a period doubling that of the number
of possible history bit-strings in the efficient regime are pointed out.

\begin{center}
{\bf ACKNOWLEDGMENTS}
\end{center}

DFZ acknowledges the support from the Natural Science Foundation of 
Guangdong Province, China.
BHW acknowledges the support from the Special Funds
for Major State Basic Research Projects in China (973 Project), 
the National Basic Research
Climbing-up Project ``Nonlinear Science'', and  the National Natural Science
Foundation in China (NNSFC) under Key Project Grant No. 19932020 and General
Project Grants Nos.19974039 and 59876039. 
We would like to thank Dr. P.M. Hui for useful discussions and for a
critical reading of the manuscript.  This work was initiated
during our visit to the Department of Physics at CUHK.
The visits were supported in part by a Grant (CUHK4129/98P)
from the Research Grants Council of the Hong Kong SAR Government.

\centerline{\bf Figure Captions}

\bigskip
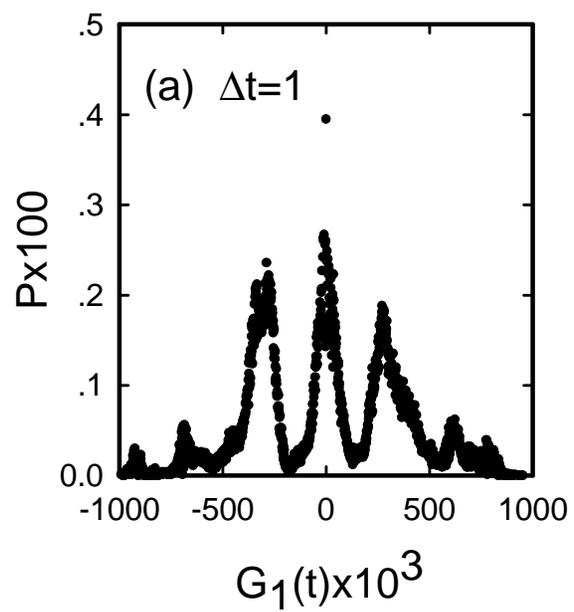
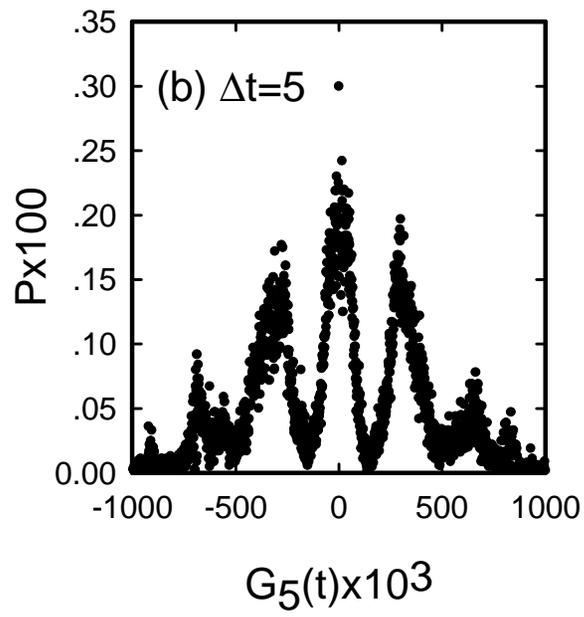
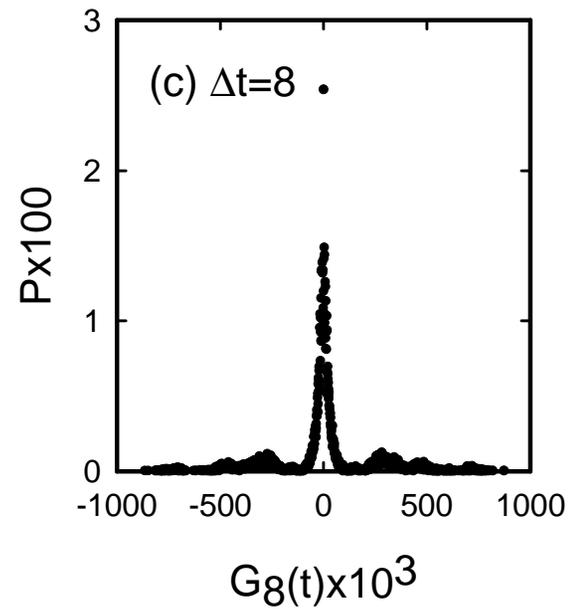
\noindent Figure 1: {Probability distribution of  $G_{\Delta t}(t)$ for $N=1001$, $m=2$ and $s=2$ with 
(a) $\Delta t =1$; (b) $\Delta t =5$; and (c) $\Delta t =8$.}

\bigskip
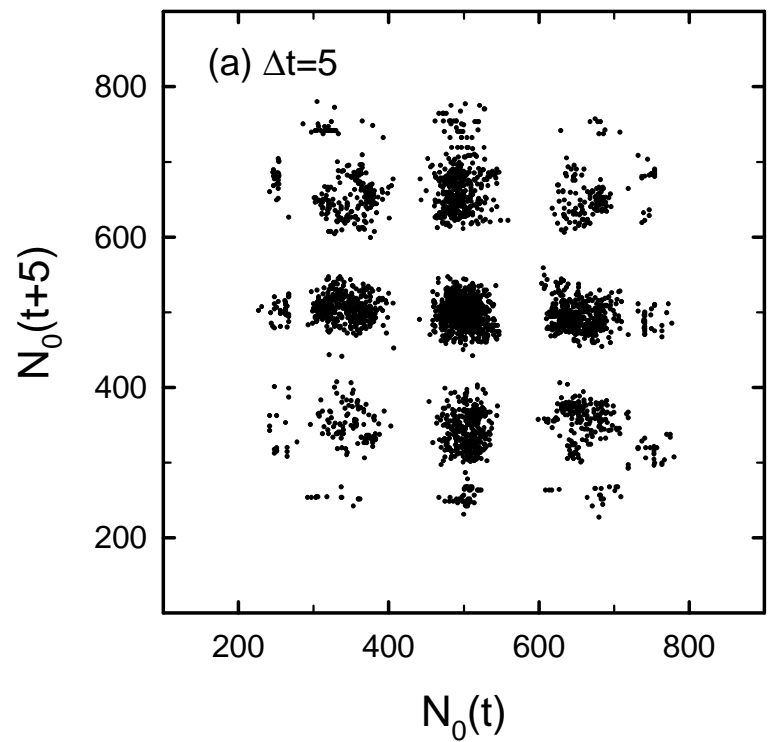
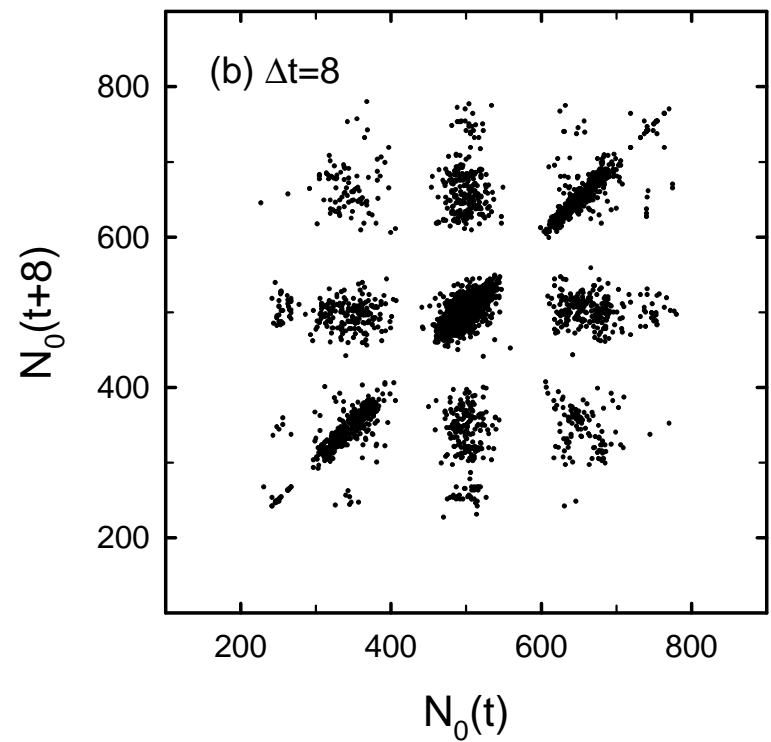
\noindent Figure 2: (a)  The plot of $N_{0}(t+5)$ against $N_{0}(t)$. (b) The plot of $%
N_{0}(t+8)$ against $N_{0}(t)$.

\bigskip
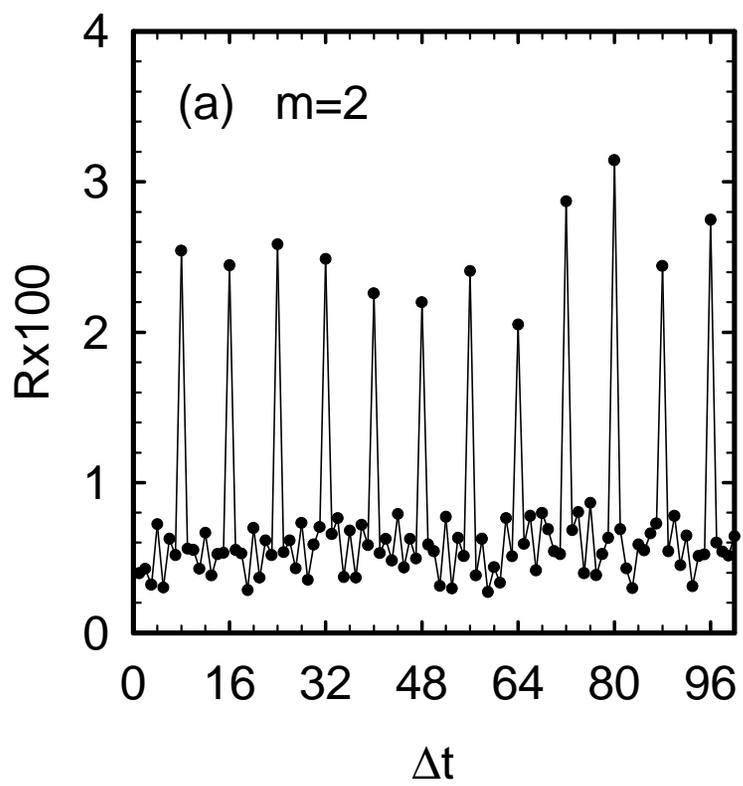
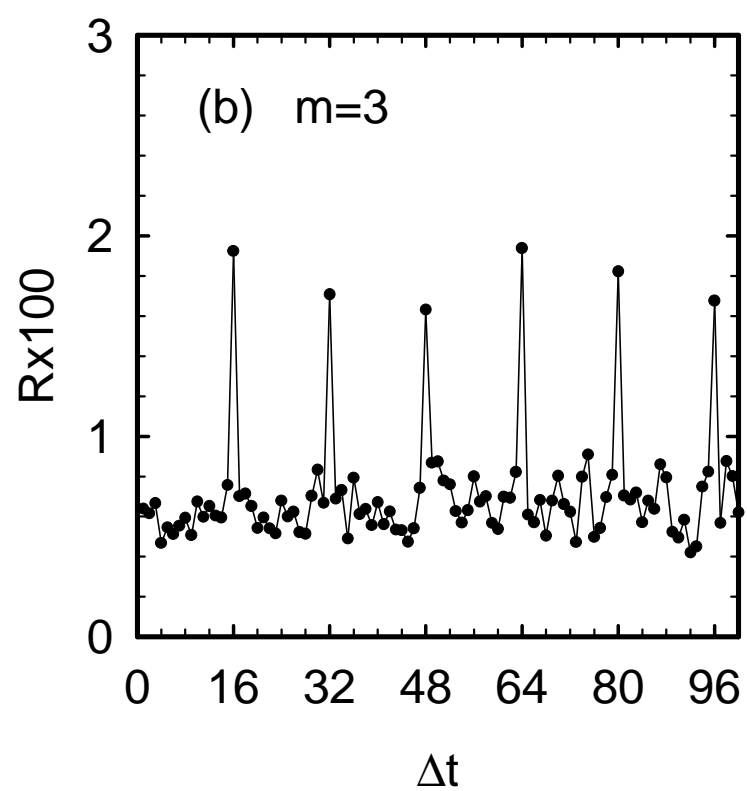
\noindent Figure 3: The probability of zero
return $R$ as a function of
$\Delta t$ for $N=1001$ and $s=2$ with (a) $m=2$ and (b) $m=3$.

\bigskip
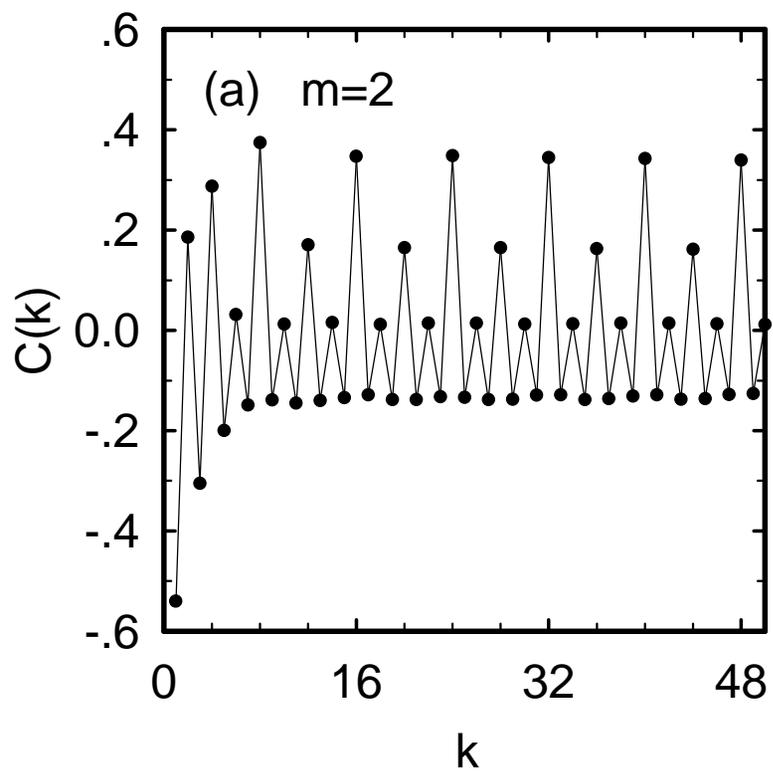
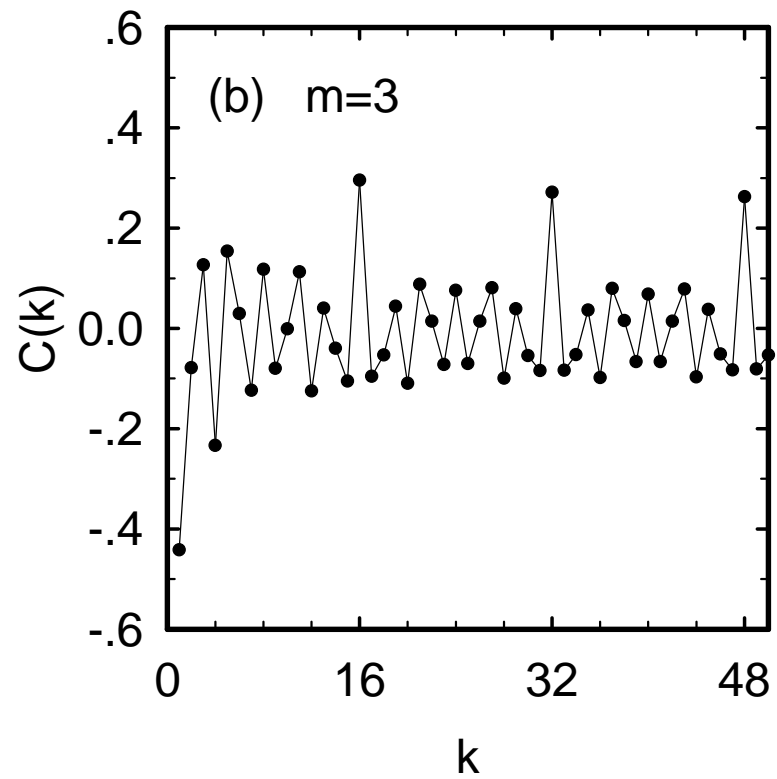
\noindent Figure 4: The autocorrelation function $C(k)$ as a function of
$k$ for $N=1001$ and $s=2$, with (a) $m=2$ and (b) $m=3$.

\end{document}